\documentclass[5p,authoryear,twocolumn]{elsarticle}

\usepackage[varg]{txfonts}
\usepackage{graphicx}
\usepackage{amssymb}
\usepackage{xspace}
\usepackage{tabularx}
\usepackage{ccaption}

\newcommand{\lcorr}{{\tt lcorr}\xspace}
\newcommand{\fringefind}{{\tt fringefind}\xspace}
\newcommand{\mr}{\mathrm}

\begin{document}

\title{The beamformer and correlator for the Large European Array for Pulsars}

\author[astron]{R. Smits}
\ead{rjm.smits@gmail.com}
\author[astron]{C.\,G.\,Bassa}
\author[astron]{G.\,H.\,Janssen}
\author[max]{R.\,Karuppusamy}
\author[max,jodrell]{M.\,Kramer}
\author[kiaa]{K.\,J.\,Lee}
\author[max,nancay]{K.\,Liu}
\author[jodrell]{J.\,McKee}
\author[inaf]{D.\,Perrodin}
\author[jodrell]{M.\,Purver}
\author[api,astron]{S.\,Sanidas}
\author[jodrell]{B.\,W.\,Stappers}
\author[max]{W.\,W.\,Zhu}

\address[astron]{ASTRON, the Netherlands Institute for Radio Astronomy, Postbus 2, 7990 AA, Dwingeloo, The Netherlands}
\address[max]{Max Planck Institut f\"ur Radioastronomie, Auf dem H\"ugel 69, 53121 Bonn, Germany}
\address[jodrell]{Jodrell Bank Centre for Astrophysics, School of Physics and Astronomy, The University of Manchester, Manchester, M13\,9PL, United Kingdom}
\address[kiaa]{KIAA, Peking University, Beijing 100871, P.R. China}
\address[nancay]{Station de Radioastronomie de Nan\c cay, Observatoire de Paris, 18330 Nan\c cay, France}
\address[inaf]{INAF - Osservatorio Astronomico di Cagliari, via della Scienza 5, 09047 Selargius (CA), Italy}
\address[api]{Anton Pannekoek Institute for Astronomy, University of Amsterdam, Science Park 904, 1098 XH Amsterdam, The Netherlands}

\begin{abstract}
The Large European Array for Pulsars combines Europe's largest radio
telescopes to form a tied-array telescope that provides high
signal-to-noise observations of millisecond pulsars (MSPs) with the
objective to increase the sensitivity of detecting low-frequency
gravitational waves. As part of this endeavor we have developed a
software correlator and beamformer which enables the formation of a
tied-array beam from the raw voltages from each of telescopes. We
explain the concepts and techniques involved in the process of adding
the raw voltages coherently. We further present the software
processing pipeline that is specifically designed to deal with data
from widely spaced, inhomogeneous radio telescopes and describe the
steps involved in preparing, correlating and creating the tied-array
beam. This includes polarization calibration, bandpass correction,
frequency dependent phase correction, interference mitigation and
pulsar gating. A link is provided where the software can be
 obtained.
\end{abstract}


\maketitle

\section{Introduction}
One of the remarkable predictions from the theory of general
relativity is the existence of ripples in space-time, called
gravitational waves (GWs), which are created for example by the
acceleration of masses. The first proof of their existence came from
the observed decay of the orbital period in compact systems of two
orbiting stars as the GWs carry energy away
\citep[e.g.][]{Taylor}. More recently, the LIGO Scientific
Collaboration and the Virgo Collaboration observed the first direct
detection of a transient GW signal originating from a binary black
hole merger~\citep{GW}. LIGO and similar detectors probe GWs at
kHz-frequencies making them most sensitive to signals from merging
binary neutron stars and black hole systems. Regular observations of
radio pulsars have the potential to probe GW frequencies down to
nanohertz. This could provide a direct detection of the stochastic GW
background originating from the ensemble of coalescing supermassive
black hole binaries throughout universe \citep[e.g.][]{haehnelt,
  jaffe, sesana} and possibly also a detection of single sources such
as near coalescing binary systems and cosmic strings
\citep[e.g.][]{single, cosmic, lommen}. Radio pulsars are spinning
neutron stars that emit narrow beams of radio emission along their
magnetic axes. As the beam passes across the telescope, a pulse of
radiation can be observed with the time between pulses corresponding
to the highly regular rotation of the neutron star. This makes pulsars
act like cosmic clocks. In a Pulsar Timing Array (PTA) experiment the
emission from the most stable pulsars, millisecond pulsars (MSPs), act
as the arms of a huge Galactic GW detector
\citep{Detweiler,Hellings}. The timing of pulsars for PTAs now spans
well over a decade and are typically performed on a monthly basis
\citep[e.g.][]{epta, ppta, nanograv}. This long time span is what
makes a PTA sensitive to GW frequencies down to nanohertz and makes it
complementary to the ground-based detectors.

The ongoing efforts for the detection of GWs via a PTA experiment are
pushing the precision limits of what is possible when using the
existing telescopes individually. They currently fall short in
achieving the required timing precision for a large enough sample of
pulsars to get a detection \citep{demorest, shannon, lentati}. The
precision can be improved by increasing the telescope sensitivity.

The Large European Array for Pulsars (LEAP) is an ERC-funded
experiment to combine the raw voltages from individual telescopes to
form a tied-array telescope to provide high signal-to-noise
observations of MSPs with the objective to increase the sensitivity
for a GW detection \citep{leap}. It combines the data from the
telescopes participating in the European Pulsar Timing Array
(EPTA). These telescopes are the Effelsberg telescope (EB), the Lovell
telescope at Jodrell Bank (JB), the Nan\c cay radio telescope (NRT),
the Sardinia Radio Telescope (SRT) and the Westerbork Synthesis Radio
Telescope (WSRT). Combined coherently, the effective area is
equivalent to that of a 195-m dish.

The observations are performed at L-band from $1332 - 1460$\,MHz in
monthly observing sessions of 25 hours. Pulsars are typically observed
for 45 minutes to 1 hour. Each pulsar observation is either preceded
or followed (or both) by an observation of a phase calibrator for a
few minutes. The calibrators were selected from the VLBA Calibrator
List\footnote{\tt http://www.vlba.nrao.edu/astro/calib} and are offset
by no more than about 5$^\circ$ from the pulsar position. The Nyquist
sampled timeseries from all observations are recorded to disk. These
disks are shipped to Jodrell Bank Observatory (JBO) where the baseband
data is correlated which yields the exact time-delays and
phase-offsets between each pair of telescopes. These delays and
offsets are then applied to the timeseries and the timeseries are
added together resulting in the LEAP tied-array beam.

There are several software correlators already in existence. Most
notably and versatile are Distributed FX (DiFX) \citep{difx} and SFXC
\citep{SFX}. DiFX is a correlator for Very Long Baseline
Interferometry (VLBI) that can utilize a multi-processor computing
environment by parallelizing the computations. It is widely used by
radio interferometers. It allows for pulsar binning, but does not do
beamforming. SFXC is the EVN data processor and is used to perform
global VLBI in real-time. It can perform pulsar binning and both
coherent and incoherent dedispersion. In recent years, the ability to
phase up the VLBI telescopes to create a tied-array beam has been
added. The benefits of developing an independent LEAP correlator and
beamformer are the natural integration of the data format used
\citep[i.e. the {\tt DADA} format,][]{dada}, bandwidths and available
computing hardware (see Section~\ref{sec:hardware}), the requirement
of accurate polarization calibration (see Section~\ref{sec:pol}), but
also the lack of beamforming capabilities in the existing software
correlators at the start of the LEAP project.

\citet{leap2010} first introduced LEAP. \citet{leap} provide a
detailed description of its experimental design and initial science
results. In this paper we describe in detail the techniques used in
the correlation process and the pipeline that has been developed
specifically for the purpose of processing the raw voltages from the
LEAP telescopes and creating the tied-array beam. All software that
has been produced as part of this pipeline is available for download
from {\tt http://www.epta.eu.org/aom.html}.

\section{Technique}
During an observing session, each of the LEAP telescopes observes the
same sources simultaneously. The raw voltages are Nyquist-sampled and
recorded as baseband data, thereby preserving both the amplitude and
phase information of the observed radio waves. This allows the
individual time series to be shifted in time and phase offline such
that all time series are perfectly aligned as though they were
observed with a (very large) single dish. Adding the time series after
they have been aligned in time and phase is called beamforming. The
resulting beam is called a tied-array beam.

The time and phase delays between the time series from individual
telescopes consists of four components. First, the largest delays are
due to differences in the path lengths that the signal has to travel
to reach each telescope, called the geometric delay, which corresponds
to a time delay. Second, there are differences between each
observatory's local clocks which introduces a time delay. The third
component consists of instrument-specific delays due to cables and
electronic components which introduce both a time and a phase
delay. Finally, the atmosphere (both ionosphere and troposphere)
introduces a time delay as well as a phase shift of the
radio-wavefront, which depend on the time-varying conditions of the
local atmosphere as well as the wavelengths of the radio waves.

The geometric delays can be largely corrected for by using the known
terrestrial positions of the telescopes, telescope pointing models and
celestial position of the source (calibrator or pulsar). This is
achieved using the software program CALC\footnote{CALC is part of the
  Mark-5 VLBI Analysis Software Calc/Solve~\citet{calc}. The LEAP pipeline uses CALC 9}. The
time-varying geometric delays from CALC are stored as polynomials in
CALC-files, which are read by the LEAP correlator (see
Section~\ref{sec:processing}). For LEAP we make use of a wrapper for
CALC, which is part of the software correlator
DiFX~\citep{difx}. Where known, the pulsar positions are corrected for
proper motion to the epoch of the observation. EB is used as the
geometric location of the LEAP tied-array beam, even when EB does not
participate in the observation session. This fixed reference
  point is necessary to obtain a consistent LEAP dataset for pulsar
  timing. The clock delays between the observatories are determined
from the local Global Positioning System (GPS) measurements at each
observatory. These are performed multiple times per day. For each
month the daily average of these measurements are placed on an
ftp-server\footnote{\tt ftp://vlbeer.ira.inaf} in GPS-files. As part
of the LEAP pipeline, these GPS-files are downloaded for each epoch
and read by the LEAP correlator.

\begin{figure*}
\centering
\includegraphics[width=\textwidth]{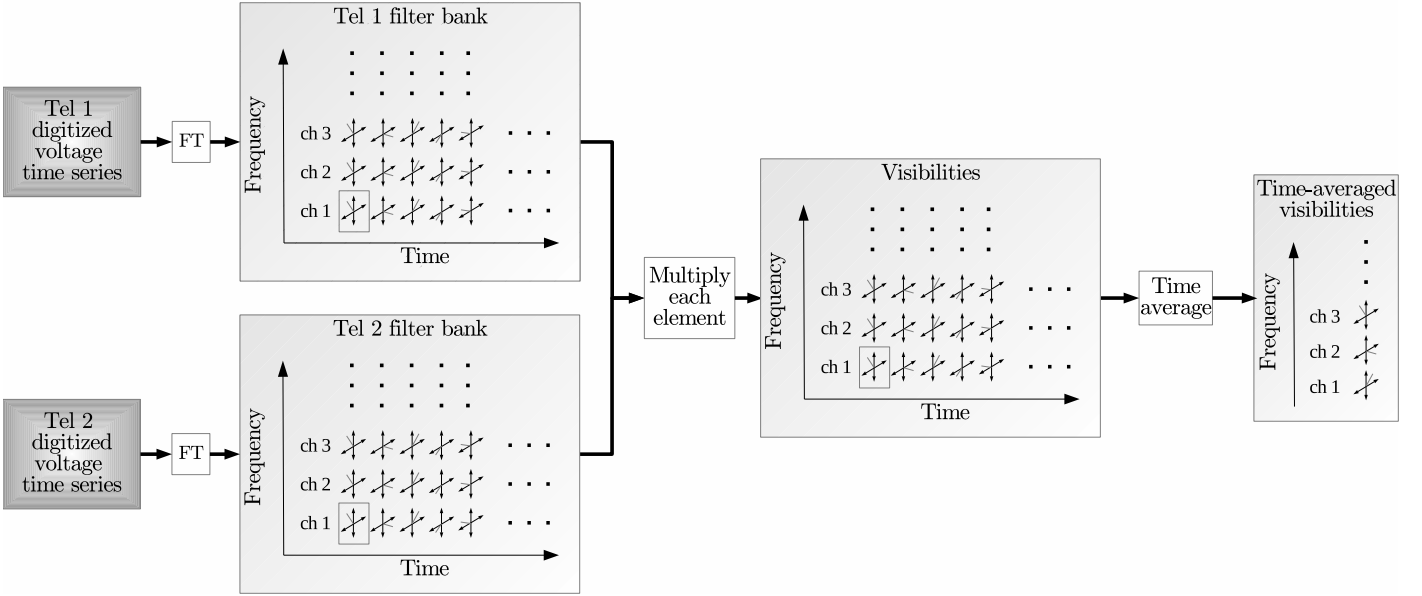}
\caption{A flowchart of the correlation process. The digitized voltage
  time series from two telescopes are Fourier transformed to form
  complex frequency channels, indicated as Tel 1 filterbank and Tel 2
  filterbank. Each element is a complex voltage that can be
  represented as a two-dimensional vector. The corresponding elements
  from both telescopes are multiplied together to form the
  visibilities. Each frequency channel is then averaged in time to
  form the time-averaged visibilities. This process is repeated for
  all telescope-pairs.}
\label{fig:correlating}
\end{figure*}

\begin{figure*}
\centering
\includegraphics[width=0.45\textwidth]{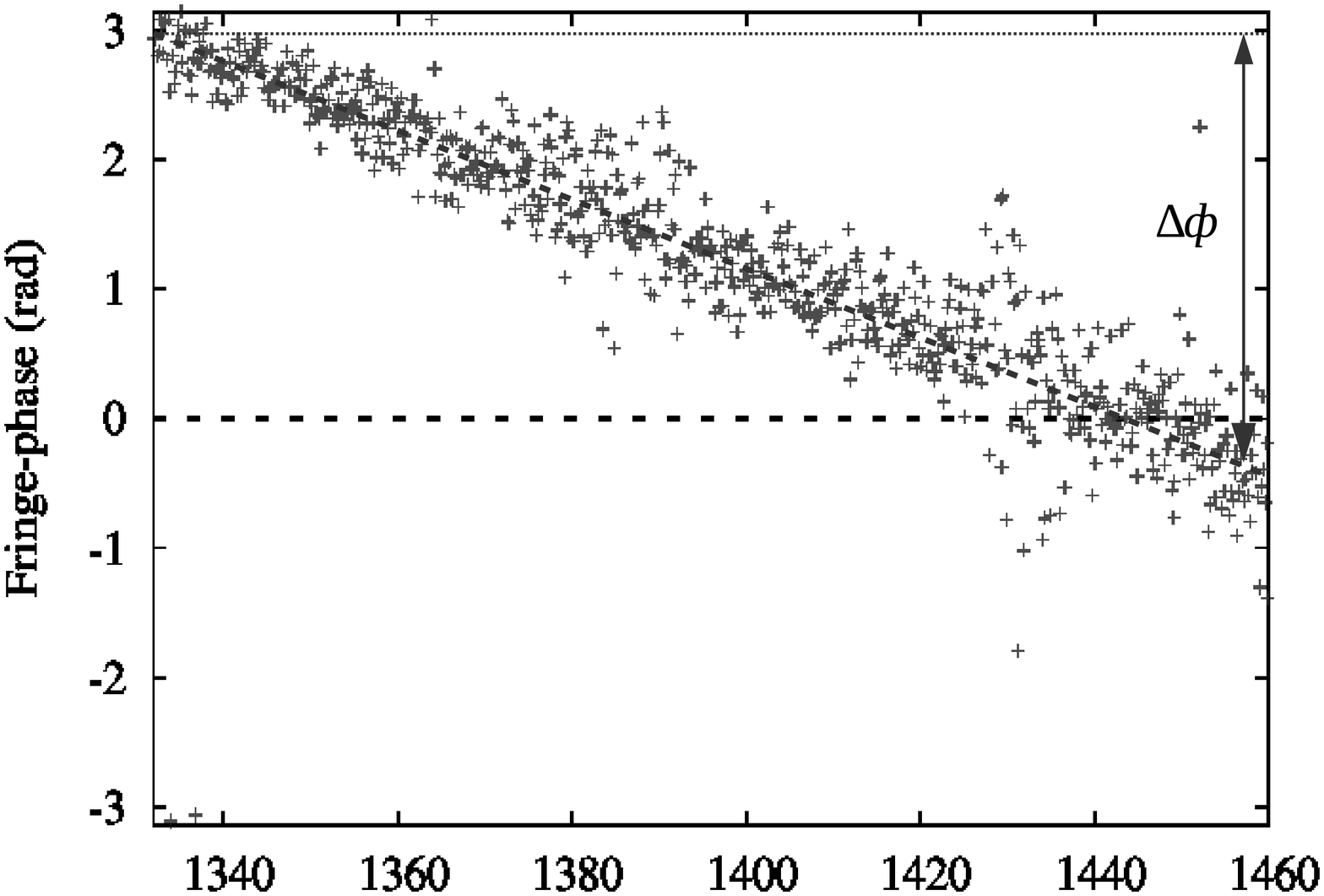}
\includegraphics[width=0.45\textwidth]{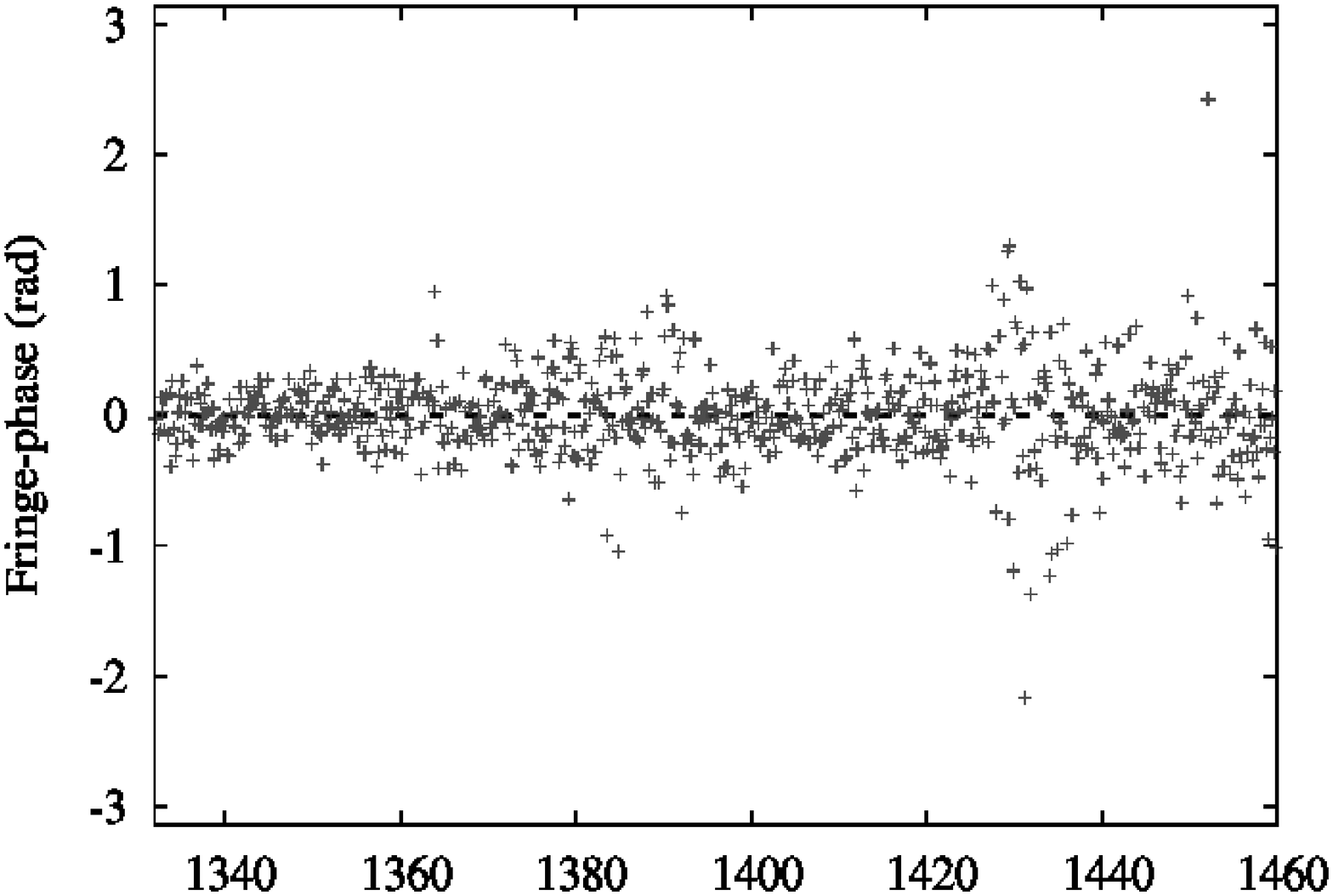}\\
\caption{Phase of the visibilities from calibrator J1025+1253 between
  telescopes EB and NRT as a function of frequency channel (after
  correcting for the frequency-dependent phase offset, see
  Section~\ref{sec:pcf}). The left panel shows a fringe between two
  telescopes that have a small time offset which introduces a
  slope. The right panel shows the fringe between two telescopes that
  are perfectly aligned, both in phase and in time. The scatter in the
  visibilities is due to the noise in the data, which does not
  correlate between the telescopes. The increase in scatter around
  1430\,MHz is due to an artifact in the EB bandpass, as can be seen
  in the left panel of Fig.~\ref{fig:bandpass}.}
\label{fig:visibilities}
\end{figure*}

The delays from the signal-path and the atmosphere are measured by
first cross-correlating the complex sampled voltages of each
polarization and each pair of telescopes from either a calibrator or
the pulsar signal itself. A cross-correlation is performed by
multiplying the Fourier transformed voltage series and integrating
over time. This procedure is visualized in Fig.~\ref{fig:correlating}.
The resulting time-averaged visibilities contain the
phase-relationship between the two time series. This can be visualised
by plotting the phase of the visibilities as a function of frequency
channel as shown in Fig.~\ref{fig:visibilities}. If they are perfectly
in phase, then the resulting visibilities all have a phase of zero. If
the phase has a constant value, then the two time series have a phase
offset, but no time offset. A time offset between the two time series
corresponds to a slope. These can be measured by applying the
fringe-find method from \citep{schwab}. This method makes use of phase
closure and involves a fit of the averaged visibilities to find the
fringe solution, which consists of the best values for the time
offsets (fringe delay), the phase offsets (fringe phase) and the phase
drift (fringe rate) (see Section~\ref{sec:fringefind}). 

Once the fringe solution is found, it is applied to the complex
samples from all telescopes which aligns them in time and phase. This
is done by rotating each complex sample by an amount $\phi[p,t]$,
which consists of:
\begin{equation}
\label{eq:phi}
\phi[p,t] = \phi^p_\mr{samples} + \phi^p_\mr{fringe\_delay} + \phi^p_\mr{fringe\_offset} + \phi^p_\mr{fringe\_rate}(t),
\end{equation}
where p is the telescope index and t is time in seconds.
$\phi^p_\mr{samples}$ corrects for the shift of the phases due to the
skipped samples as follows:
\begin{equation}
\phi^p_\mr{samples} = 2\pi \mr{N}^p_\mr{samp}\frac{f_\mr{sky}}{\mr{B}},
\end{equation}
where N$^p_\mr{samp}$ are the number of skipped samples for telescope
$p$, $f_\mr{sky}$ is the sky frequency of the band\footnote{For real
  sampled data this is the lower edge of the band. For
  complex sampled data this is the centre frequency.} and B
is the bandwidth. $\phi^p_\mr{fringe\_delay}$ is the phase from the
fractional delay for telescope p. Rotating the complex voltages by
this phase corresponds to applying the residual time-delay that is
less than one sample. This time-delay consists of the fractional
sample delay and the delay measured from the visibilities, which
consists of any residual from the predicted clock offset and the
contribution from the time-varying ionosphere. The phase from the
fractional delay is calculated as follows:
\begin{equation}
\phi^p_\mr{fringe\_delay} = 2\pi \tau^p_\mr{frac}\frac{f_\mr{band}}{\mr{B}},
\end{equation}
where $\tau^p_\mr{frac}$ is the fractional delay for telescope p,
$f_\mr{band}$ is the frequency within the downsampled band (from 0 to
B).  $\phi^p_\mr{fringe\_offset}$ is the phase offset of the
time-series from telescope p with respect to the time-series from the
reference telescope at frequency 0. This phase offset is calculated
from the fringe-solution (see Section~\ref{sec:fringefind}) by
interpolating the slope of the fringe to a frequency of
0. $\phi^p_\mr{fringe\_rate}$ is defined as:
\begin{equation}
\phi^p_\mr{fringe\_rate}(t) = t\cdot r^p_\mr{fringe},
\end{equation}
where $t$ is the time in seconds from the start of the fringe-solution
and $r^p_\mr{fringe}$ is the fringe rate in radians per second for
telescope p, which is also a direct product from the fringe-fitting.

\section{Storage, transport and processing hardware}
\label{sec:hardware}
The baseband data from every monthly 25-hour session are initially
stored at each observatory locally \citep[the pulsar hardware at EB,
  JBO, SRT and WSRT are described in][]{puma2, roach}. For EB, JBO, NRT
and SRT the data are stored in 8 subbands of 16\,MHz with the center
frequencies at 1340, 1356, 1372, 1388, 1404, 1420, 1436 and 1452\,MHz
for a total bandwidth of 128\,MHz. The hardware present at the WSRT
generates 8 $\times$ 20\,MHz subbands. These subbands are thus
arranged to overlap with the 8 subbands from the other
observatories. At all observatories both polarizations are recorded at
8 bit resolution in the {\tt DADA} format, where each file contains 10
seconds of one subband. Table~\ref{tab:parameters} shows all the
parameters of the recorded baseband data for each of the
observatories, as well as for the resulting LEAP baseband data.

\begin{table}
  \caption{The parameters for the baseband data for each of the 5 LEAP
    observatories as well as for the LEAP data.}
  \begin{tabular}{llll}
    \hline
    \hline
    Telescope & Bandwidth of & Polarization & Nyquist\\
    & subbands (MHz) & & sampling\\
    \hline
    EB & 16 & Circular & complex \\
    JB & 16 & Circular & complex \\
    NRT & 16 & Linear & complex \\
    SRT & 16 & Linear & complex \\
    WSRT & 20 & Linear & real \\
    LEAP & 16 & Circular & real \\
    \hline
  \end{tabular}
  \label{tab:parameters}
\end{table}

The data volume from one observing session is about 36\,TB for EB, JB
and SRT. For NRT it is about 22\,TB because it participates in fewer
sources. For WSRT it is about 50\,TB due to its larger subbands. These
data are stored on 16 times 3 or 4\,TB disks.

The baseband data from each site are transported to the JBO either via
shipment of the disks, or over the internet for offline
processing. The central storage at JBO can hold over two months of
unprocessed LEAP data from all the sites, keeping it directly
accessible for correlating.

The processing is performed on a high performance computer cluster
located at JBO. This cluster consists of 40 nodes, each having two
Quad core Intel Xeon processors, 4GB of RAM and 2TB of storage.

\section{Processing Pipeline}
\label{sec:processing}
Once the baseband timeseries from a new observation are available at
the central storage, the LEAP pipeline is set up to allow multiple
users to process any part of the monthly observations. All files are
indexed and their location and relevant header information are stored
in a file-location file (FLF), which is made available to all
users. The clock correction files are downloaded and are also made
available to all users. A polarization calibration procedure is
performed on a calibration pulsar which yields a 4$\times$4 matrix,
called a Mueller matrix, for each telescope. From these Mueller
matrices, the Jones matrices are produced for each individual
observation and each telescope (see Section~\ref{sec:pol}). Based on
the FLF, a directory structure is created for each user where the FLF
is split up into smaller files containing all the relevant information
for each individual observation. The CALC files are created for each
observation, as well as input files that contain telescope and
observation specific information about the correlation or
addition. These input files can be modified by software later in the
pipeline, or by the user directly.

\begin{figure*}
\centering
\includegraphics[width=\textwidth]{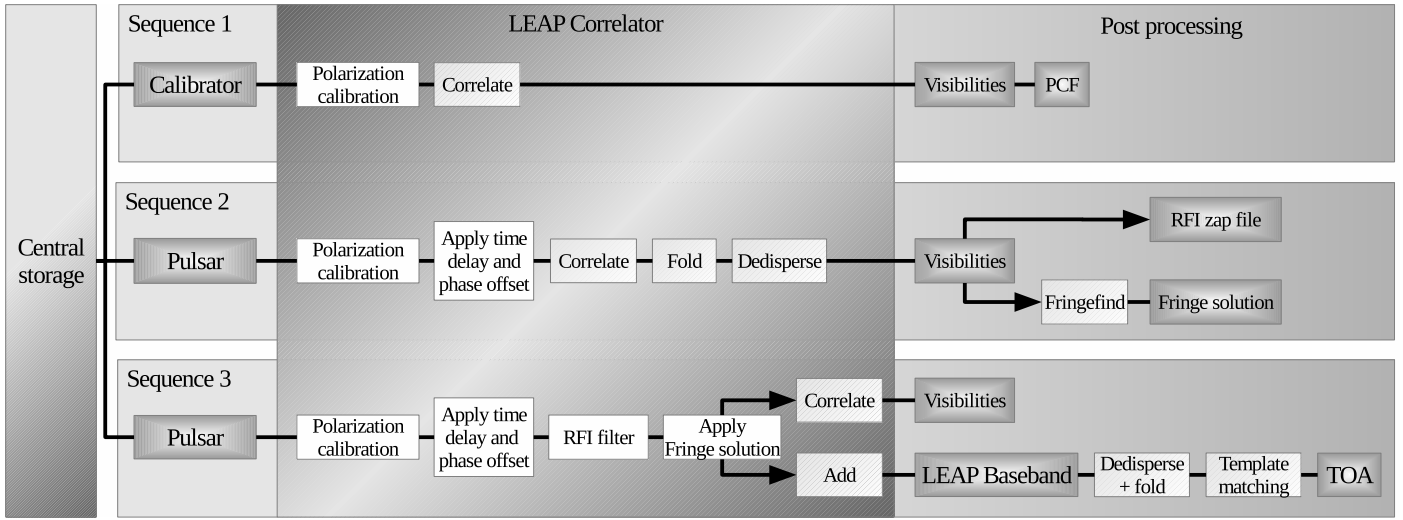}
\caption{A flowchart of the LEAP pipeline. Each observatory stores the
  raw voltages from the LEAP observations on disk. The data are then
  transferred to the central storage (at Jodrell Bank). From the
  central storage the data is correlated after applying polarization
  calibration and RFI mitigation filters. The resulting
  fringe-solutions from each baseline are then applied to the
  voltages, again after applying polarization calibration and RFI
  mitigation, and they are added together in phase, resulting in the
  LEAP phased array. This stage also includes correlating the voltages
  after the fringe-solution is applied. The resulting visibilities are
  checked to verify that the applied fringe-solution is indeed
  correct. The added voltages are processed as normal timing
  data. They are dedispersed using {\tt DSPSR} and template matching
  is performed to produce the final pulse times-of-arrival. }
\label{fig:flowchart}
\end{figure*}

At this stage the correlations and additions can be performed to form
the LEAP baseband time series, containing the tied-array
beam. Creating this time series takes place on the computer cluster
and requires several computational steps. Figure~\ref{fig:flowchart}
shows the flowchart of the LEAP pipeline. There are three sequences,
corresponding to an initial calibration of the frequency-dependent
phase offsets between all telescope pairs (sequence 1), the
correlation of the pulsar signal resulting in the fringe solution
(sequence 2) and the application of the fringe solution to the complex
samples and formation of the LEAP tied-array beam (sequence 3). Each
sequence consists of copying the required {\tt DADA} files from the
central storage to the computer cluster. The files are read and
processed by the LEAP correlator, called \lcorr (see
Section~\ref{sec:lcorr}).

In sequence 1 the voltages from a calibrator source are polarization
calibrated and the geometric delay and clock offsets are applied. The
voltages are then correlated, yielding visibilities from each pair of
telescopes, which are used to measure the frequency-dependent phase
offsets between all telescope pairs (see Section~\ref{sec:pcf}). These
offsets are written to disk in phase correction files (PCFs) and later
applied to the raw voltages of the pulsar in sequence 2 and
3. Sequence 1 only needs to be performed once for the entire
observation session.

In sequence 2 the voltages from the pulsar are polarization calibrated
and the time and phase offsets from the geometric delays, the clock
offsets and the frequency-dependent phase offsets are applied. The
voltages are then correlated, folded with the pulsar period and
dedispersed to produce the visibilities for each pair of telescopes
for each frequency channel and each pulsar phase-bin. The visibilities
in the phase-bins with the pulsar signal are averaged together (see
Section~\ref{sec:gating}) and are then processed via the algorithm as
described by~\citet{schwab}. This yields the phase and time offset and
phase drift between the pairs of telescopes, called the
fringe-solution (see Section~\ref{sec:fringefind}). Because of clock
drifts and the changing ionosphere and troposphere a new fringe-solution needs to be
found every five to ten minutes. Thus, this sequence is performed in
several independent parts, with each part yielding a new
fringe-solution. The powerspectrum of the visibilities are searched
for outliers, which are marked as radio frequency interference
(RFI). Those channels will be 'zapped' when adding the data in
sequence 3.

Sequence 3 is the beamformer. The voltages from the pulsar are again
polarization calibrated and the time and phase offsets from the
geometric delays, the clock offsets and the frequency-dependent phase
offsets are applied. The frequency channels which are marked as RFI
are masked by replacing the content with Gaussian noise with mean and
rms determined from neighboring time samples (see
Section~\ref{sec:rfi}). The fringe solution is applied. The voltages
are then added together, forming the tied array beam, and are written
to disk as the LEAP baseband data. The baseband data is dedispersed
using {\tt DSPSR} \citep{dspsr}. Template matching is performed to
produce the final pulse times-of-arrival (TOAs) \citep{template}. The
LEAP baseband data is stored on a long-term archive. In parallel to
adding the voltages, the voltages are also correlated as in sequence
2, so as to produce the visibilities with the exact same fringe
solution as the tied array beam.

All steps are performed by user interaction with code written in C,
python and bash scripts.

\subsection{The LEAP correlator}
\label{sec:lcorr}
At the heart of the LEAP correlations lies the C-based program \lcorr
which reads the individual files containing the digitized voltage time
series from each telescope, applies the time and phase corrections,
polarization calibration, RFI mitigation and can either correlate the
time series to produce the visibilities (mode 1), add them together to
produce the tied-array (mode 2), or write out the complex samples for
each telescope (mode 3). 

\begin{figure*}
\centering
\includegraphics[width=\textwidth]{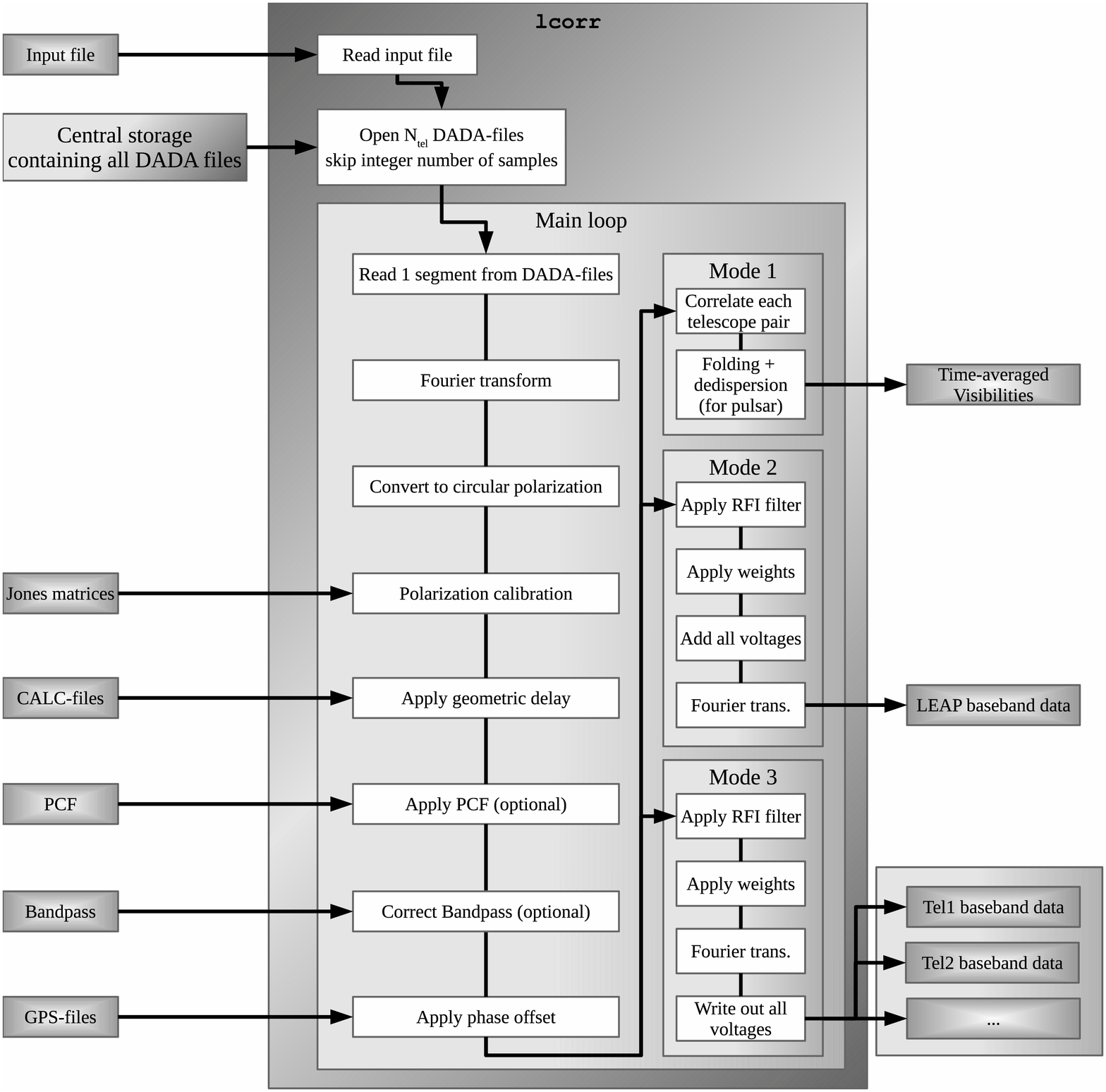}
\caption{Flowchart of the LEAP software correlator \lcorr. This chart
  corresponds to the `LEAP Correlator' block in
  Fig.~\ref{fig:flowchart} but with more details. Mode 1 is the
  correlator mode that is used in sequence 1 and 2 from
  Fig.~\ref{fig:flowchart}. Mode 2 is the addition mode that is used
  in sequence 3 from Fig.~\ref{fig:flowchart}. Mode 3 is an extra mode
  in which the complex samples are written to disk which allows them
  to be added incoherently. See text for a detailed explanation of
  this chart.}
\label{fig:lcorr}
\end{figure*}

Figure~\ref{fig:lcorr} shows the flowchart of \lcorr, which describes
the `LEAP Correlator' block in Fig.~\ref{fig:flowchart}. The program
runs on ten seconds of data from each telescope for one frequency band
at a time. It first reads the input file which contains the input
parameters for the correlation or addition, which includes the phase
and time offset for each telescope if these have been obtained from an
earlier correlation. \lcorr also reads the files containing the Jones
matrices, the CALC files, the frequency-dependent phase offsets, the
bandpass and the GPS clockfiles. The Jones matrices are used for
polarization calibration. The CALC files contain the polynomials from
which the exact geometric delays are calculated for each observatory
with EB as the reference point. The frequency-dependent phase offsets
are stored in the PCF (see Section~\ref{sec:pcf}). After reading the
GPS clockfiles, \lcorr performs a linear interpolation between the
measured clock offsets to estimate the clock offset corresponding to
the time of the ten-second time series. It then opens the N$_\mr{tel}$
{\tt DADA} files for reading, where N$_\mr{tel}$ is the number of
telescopes, and skips a number of samples corresponding to the total
time offset. This leaves a time offset less than a fraction of a
sample. Note that by skipping these samples, the phases of the
radiowaves as received by the telescopes are shifted as well. This is
corrected for at a later stage (see Section~\ref{sec:delays}).

Within the main loop, \lcorr reads the content of the {\tt DADA} files
for each telescope in small segments, typically 100
samples\footnote{Typically 100 samples are used for the Fourier
  transform of 16-MHz data and 125 samples are used for the Fourier
  transform of 20-MHz data, producing a filterbank with 160-kHz
  channels. This ensures that the channels from both the 16-MHz data
  and the 20-MHz data from the WSRT can be matched in frequency.}. The
samples are Fourier transformed. If the polarization of the data is
linear, it is converted to circular so that the polarization from all
telescopes are in the same basis. This is performed using
the conversion Jones matrix as described in~\citet{hamaker}. The data
is then polarization calibrated by applying the Jones matrix for that
telescope (see Section~\ref{sec:pol}) and the geometric delay is
applied. Optionally, the frequency-dependent phase offsets are applied
(in practice, these are always applied once they have been
measured). Optionally, a bandpass correction is applied (see
Section~\ref{sec:bandpass}). Next, the segments from all telescopes
are aligned in phase. This happens by rotating each complex sample by
an angle $\phi$, which aligns the signal within the samples from all
telescopes in phase (see Section~\ref{sec:delays}). \lcorr can then
either multiply the voltages between each pair of telescopes,
resulting in the visibilities (mode 1), add the voltages together
(mode 2), or write out the complex samples for each telescope (mode
3).

If the time series are from a pulsar observation, then in mode 1 the
visibilities from all frequency channels are binned into a given
number of phase-bins over the pulsar period. The pulsar period during
the 10-second time series is predicted using {\tt TEMPO}, a program
for analyzing pulsar timing data~\citep{tempo} and the pulsar
parameter file (i.e. the $<$psrname$>$.par file). The binned
visibilities are dedispersed incoherently by shifting the frequency
channels by the appropriate integer number of phase-bins using a
cyclic boundary. For a calibrator, the visibilities from each
frequency channel are simply averaged over time. In mode 2, the
samples are first scaled according to a weight defined in the input
file. This will correct for differences in telescope gain and optimize
the S/N (see Section~\ref{sec:adding}). The samples are then added
together and scaled to a dynamic range of 8 bits before being written
to disk in the {\tt DADA} file format. Mode 3 is identical to mode 2,
except that the complex samples from the telescopes are not added
together, but are written to disk individually in the {\tt DADA} file
format which allows them to be added incoherently (see
Section~\ref{sec:align}).

\subsection{Polarization calibration}
\label{sec:pol}
A crucial part of the correlator pipeline, is an accurate
polarization calibration that removes the effects introduced by the
telescope, receiver and instruments.  For this purpose we observe a
suitable polarization calibrator with all participating telescopes as
part of the observation session. This is typically a bright, highly
polarized pulsar such as PSR~B1933$+$16, or PSR~J1022$+$1001. After
dedispersion, the frequency channels are averaged together so as to
yield a frequency resolution of 4\,MHz, which provides both sufficient
sensitivity as well as sufficient frequency resolution. Then, for each
4-MHz channel of the pulsar observation from each telescope, a
polarization calibration procedure is performed which yields the
4$\times$4 Mueller matrices. These Mueller matrices describe the
distortion of polarization and are independent of the paralactic
angle. From the Mueller matrices the Jones matrices can be produced,
which describe the distortion of polarization for a specific
paralactic angle. Since the paralactic angle changes between sources
(for the alt-azimuth mounted telescopes), there is a Jones matrix for
each 4-MHz channel from each telescope and for each observation. The
changes in telescope angle on the sky during an observation is usually
neglected.

The Jones matrices are read by \lcorr. Once the complex samples from
each telescope are Fourier transformed and converted to circular
polarization, the samples are multiplied by the appropriate Jones
matrix, which yields the polarization calibrated complex samples. This
process is performed before the addition to maximize the coherency of
the tied-array beam, but also because performing the polarization
calibration after addition is complicated due to the extra phase
rotation that is applied to the complex samples of each telescope.
~\citet{leap} give a description of the LEAP polarization calibration,
which is a crucial step to achieve full coherence. For this
  reason, a more detailed description can be found in a forthcoming
paper (Lee et al., in prep.).

\subsection{Bandpass correction}
\label{sec:bandpass}
\begin{figure*}
\centering
\includegraphics[width=0.45\textwidth]{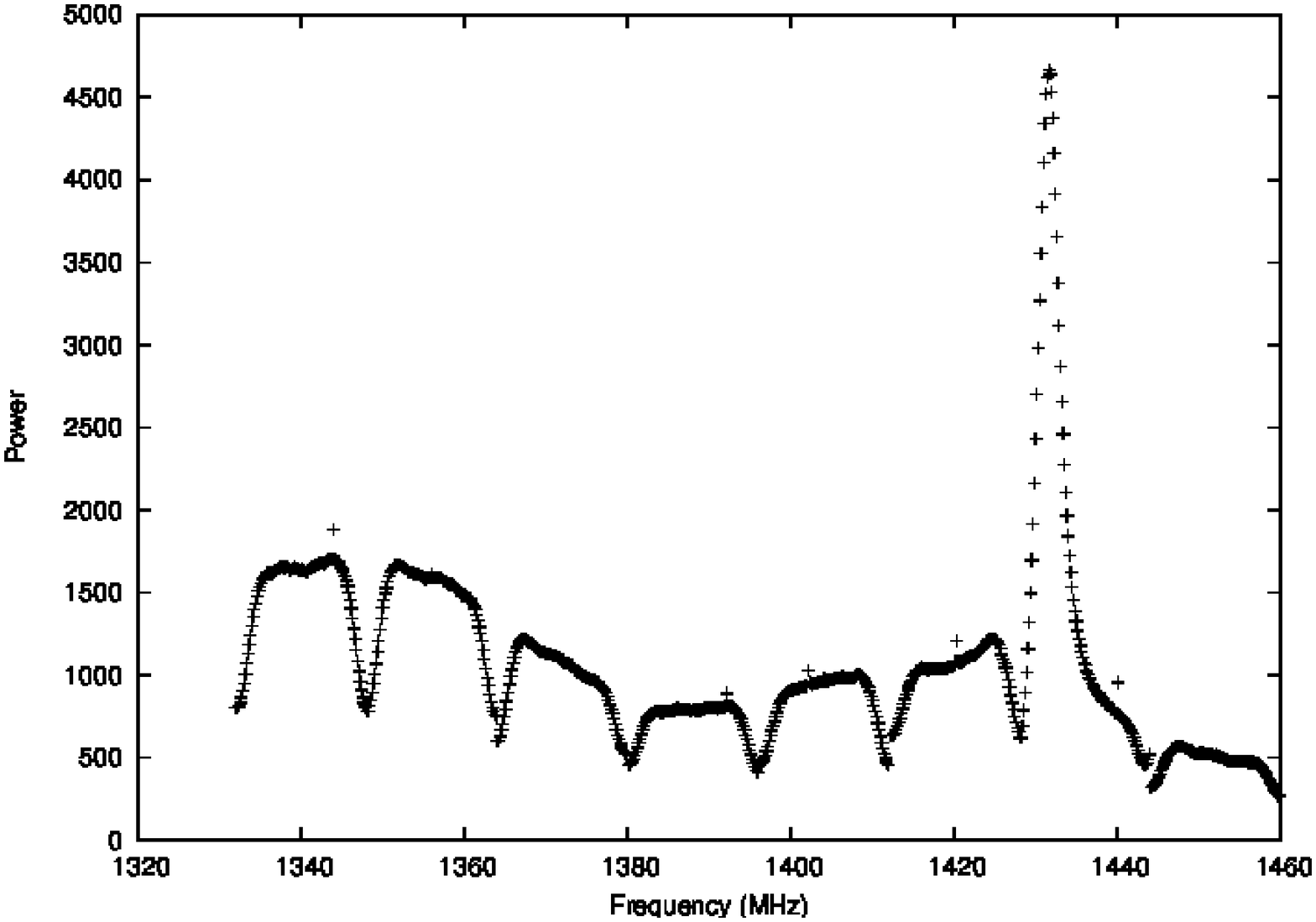}
\includegraphics[width=0.45\textwidth]{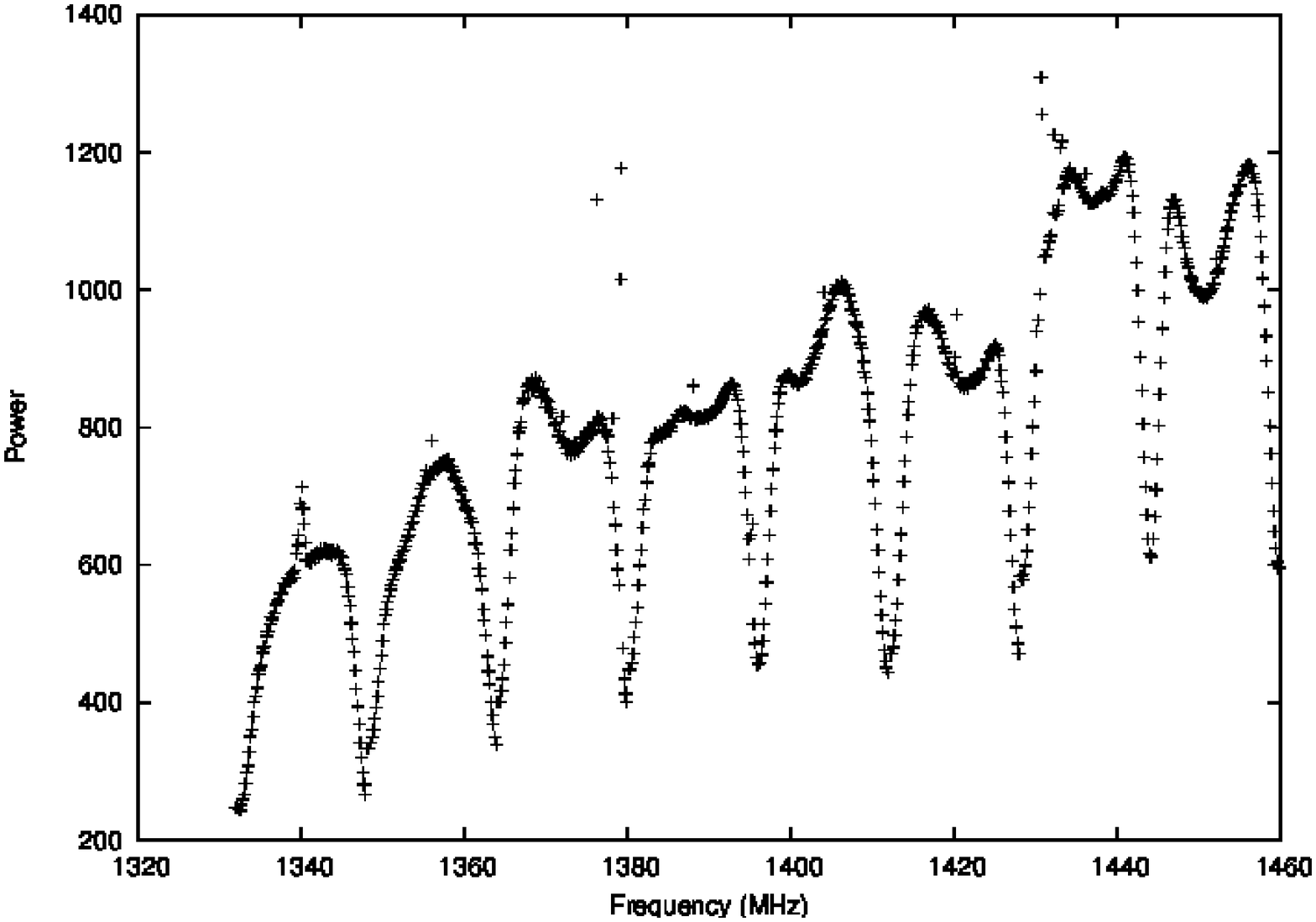}
\caption{The bandpass of 8 16-MHz bands from EB (left) and NRT
  (right). After polyphase filtering, the EB bandpass shows an
  artificial feature around 1430\,MHz caused by a resonance in the
  receiver. Some channels have a clear excess of integrated power, which
  is caused by narrow-band RFI. Such channels will be masked by the
  interference mitigation (see Section~\ref{sec:rfi}).}
\label{fig:bandpass}
\end{figure*}
The receiving power of an antenna within a band is affected by the
receiver antenna and the polyphase filter. For a perfect system, this
bandpass would be flat. In practice, however, the bandpass usually has
a roll-off at the edges of the band and may show other features. See
for example Fig.~\ref{fig:bandpass} which shows the bandpass from both
EB and NRT. These features have a small negative effect on the
addition of the signal from different telescopes as they introduce an
artificial weight on each frequency channel. The LEAP software allows
for a bandpass correction. This is performed by first correlating a
bright calibrator that has been observed right before the pulsar
observation. From the visibilities of the autocorrelations, we extract
the average receiving power of each telescope for both polarizations
for each frequency channel, to obtain the bandpass. These values are
stored in data files, which can be read by \lcorr. The option for a
bandpass correction can be activated via the input file. \lcorr will
then divide the complex samples in each frequency channel with the
corresponding average receiving power, thus normalizing the bandpass.

\subsection{Frequency dependent phase correction}
\label{sec:pcf}
\begin{figure*}
\centering
\includegraphics[width=0.45\textwidth]{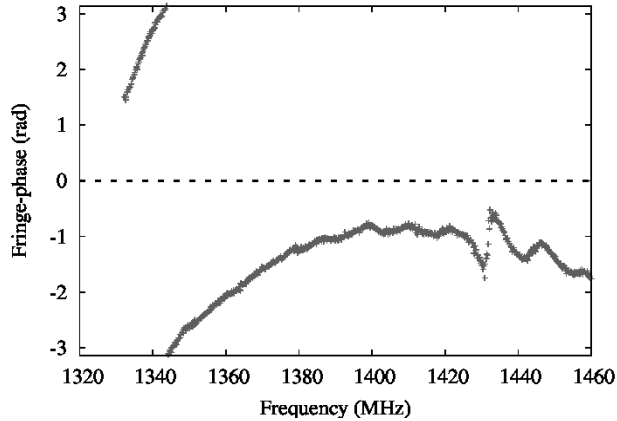}
\caption{Phase of the visibilities from calibrator J1025+1253 between
  telescopes EB and NRT as a function of frequency channel before
  correcting for the frequency-dependent phase offset.}
\label{fig:pcf}
\end{figure*}
The receiver antenna and the polyphase filter not only affect the
power of an antenna, but also the phase of the complex sampled
voltages. However, the frequency evolution of the signal's phase only
becomes visible when a reference signal is present, as is the case
when correlating the baseband data from 2
telescopes. Fig.~\ref{fig:pcf} shows the phase of the visibilities
from a calibrator between two telescopes before correcting for the
frequency evolution of the signal's phase.  To keep both signals in
phase over the whole band, the frequency-dependent phase offset needs
to be measured. This is performed by first correlating a bright
calibrator that has been observed with all the telescopes. This can be
the same correlation as used for the bandpass correction. From the
visibilities of each pair of telescopes the phase-offset is calculated
for each frequency channel and for both polarizations. These
phase-offsets are stored in the PCFs, which can be read by \lcorr. The
option for a phase correction can be activated via the input
file. \lcorr will then rotate the complex sample of every visibility
by the corresponding phase, removing all frequency dependence of the
phases.

The combination of correcting for both the power and the phase
in the bandbass is sometimes referred to as a complex bandpass
correction.

\subsection{Interference mitigation}
\label{sec:rfi}
The method for interference mitigation is described in Section 4.5 of
~\citet{leap}. In brief, there are two techniques that are both
implemented within \lcorr. 

The first technique involves selecting frequency channels from the
Fourier transformed voltages from each telescope that contain an
excess of integrated power. This technique is sensitive to narrow-band
interference. Channels are selected when the power exceeds a threshold
or deviates significantly from its neighbors. These channels are then
masked by replacing the content with Gaussian noise with a mean and
rms determined by the average of the remaining frequency channels.

The second technique implements the method of spectral kurtosis and is
most sensitive to time-varying interference
\citep[see][]{nita1,nita2}. In each frequency channel and at each
telescope, the distribution of a time-series of 1000 samples of total
power is assessed for similarity to that expected from
Gaussian-distributed amplitudes. Non-Gaussian distributed amplitudes
are attributed to the presence of interference. The samples from such
a time-series are replaced with Gaussian noise.

\subsection{Pulsar gating}
\label{sec:gating}
Mode 1 of the correlator \lcorr produces the visibilities from all the
telescope pairs. These visibilities are then used to find the
fringe-solution, which is required to align the timeseries in time and
phase (see Section~\ref{sec:fringefind}). The observation of a
calibrator is used to find an approximate fringe-solution for the
pulsar observation. Due to the temporal and spatial offset of the
calibrator that solution will deviate from the required
fringe-solution for the pulsar observation. Also, during the course of
the pulsar observation any fixed fringe-solution will become invalid
due to clock drifts and the changing ionosphere and troposphere. It is thus required
to calibrate on the pulsar signal itself. The processing of the pulsar
timeseries is performed in several parts to allow for different
fringe-solutions over the course of the observation. These parts are
typically five to ten minutes long. The resulting visibility files
contain the visibilities from all the telescope pairs in a given
number of phase-bins, where all the phase-bins make up one pulse
period. Typically, the number of phase-bins is about 30. For each of
these phase-bins the average power is calculated and the pulsar
profile is created. Phase-bins that have a power of more than 3 sigma
above the noise-level are selected as the on-pulse. This process is
called `gating'. The visibilities of the on-pulse bins are then
averaged together, which improves the S/N. The resulting visibilities
are used to find a fringe-solution.

\subsection{Finding fringes}
\label{sec:fringefind}
The correlations are run in parts, where each part contains all the
bands and all the 10-second time-series over a given time-interval of
typically five to ten minutes. After the correlations from one part
are finished, all resulting visibilities are merged together to form
one visibility file. For a pulsar observation, the visibilities are
gated (see Section~\ref{sec:gating}). The merged visibility file is
used to find the fringe-solution.  The fringe-search is performed via
the C-based program \fringefind, which applies the method of
~\citet{schwab} in two stages. In the first stage the visibilities are
Fourier transformed and each resulting complex sample is multiplied by
its complex conjugate to yield a power series. From this series, the
sample with maximum is power is selected, which corresponds to a delay
which is a multiple of the Nyquist sampling of the original voltages
(i.e. 62.5\,ns for the 16-MHz bands). In the second stage this integer
shift is applied to the visibilities. A least-squares algorithm is
used that minimizes the difference between model phases and measured
phases by solving for the fringe phase, the fringe delay and the
fringe rate between each pair of telescopes. The resulting solution is
restricted by the requirement of phase closure. The fits are performed
independently on both left-hand-circular and right-hand-circular
polarizations. The resulting fringe rates are averaged over both
polarizations. Typically the initial fringe-solution is found using a
calibrator. This solution is then applied to the pulsar observation,
which places the fringe-delays between each pairs of telescopes
already within one Nyquist sample.

\subsection{Correcting the time delays}
\label{sec:delays}
Before the addition, the complex samples from all telescopes are
aligned in phase. This happens by rotating each complex sample by an
amount $\phi[k,p,t]$, as defined in Eq.~\ref{eq:phi} and obtained via
the procedure described in Section~\ref{sec:fringefind}.  After the
complex samples from each telescope are rotated by the angle $\phi$,
the contributions from the pulsar (or calibrator) signal are in phase,
whereas the contribution from the noise remain with a random
phase-offset.

\subsection{Coherent addition of the voltages}
\label{sec:adding}
Once the complex samples from each pair of telescopes are in phase,
they can be added with \lcorr (mode 2) to form the tied array beam. As
described in Section~4.4 of ~\citet{leap} we have to apply an
appropriate weight to the baseband data from each of the telescopes,
to ensure maximum S/N of the added data. The program \lcorr measures
the average noise levels of both polarizations for each telescope from
the baseband data, which is used to normalize all samples to an equal
noise level. It then applies a weight based on the S/N of the average
intensity profiles from the individual telescopes, given by:
\begin{equation}
  W_{\mathrm{tel}}=\sqrt{\frac{\mr{S/N_{tel}}}{\mr{S/N_{ref}}}},
\end{equation}
where $\mr{S/N_{tel}}$ is the S/N of the telescope and
$\mr{S/N_{ref}}$ is the S/N of a reference telescope. Finally, the
samples are again normalized such that the resulting added samples
have a proper dynamic range, yet minimum saturation when they are
converted to 8 bit. This normalization factor is given by
$35/\Sigma_{i=1}^\mr{N_{tel}} W_i$, where 35 is the rms of the added
samples and is chosen such that less than 0.3\% of the samples are
clipped.

Once the complex samples are properly scaled, they are added together,
Fourier transformed back into the time domain and written to disk as
8-bit samples in the {\tt DADA} format.

\subsection{Incoherent addition of the voltages}
\label{sec:align}
The coherent addition of the pulsar signal from different telescopes
requires a fringe solution which is obtained from correlating the
complex sampled voltages from each telescope. However, some
observations only have a very weak pulsar signal which can fail to
yield a fringe solution. In that case it is still beneficial to add
the samples together incoherently. This is achieved by using mode 3 of
\lcorr which aligns the complex samples from each telescope by
applying the geometric delays, the clock offset, the instrument
specific delays and the fringe solution from the calibrator. This
aligns the samples from the different telescopes to about 10\,ns,
depending on the conditions of the ionosphere and the calibrator
angular separation from the pulsar. The same weights as described in
Section~\ref{sec:adding} are applied, the frequency channels from each
telescope are Fourier transformed back to the time domain and are
written to disk as 8-bit samples. Each time series is then dedispersed
using {\tt DSPSR}. The resulting archive files are added with {\tt
  PSRADD}, which is part of the {\tt PSRCHIVE} software
\citep{psrchive}.

\section{Summary}
As part of LEAP, we have developed a software correlator and
beamformer that allows the pulsar observations from the telescopes
from the EPTA to be added coherently, thus significantly improving
their effective area to that of a steerable 195-m circular
dish. Another important improvement is the accurate polarization
calibration that is performed before the individual time-series are
added together. In this paper we have presented a detailed overview of
the software pipeline and the techniques that are used. In particular
it describes the steps that are required to calibrate, correlate and
add the complex sampled time series from the EPTA telescopes. It also
gives a detailed outline of the C-based program \lcorr, which lies at
the heart of the correlator and beamformer. The majority of the
software and techniques can be applied to other telescopes as well.
All software is available for download from {\tt
  http://www.epta.eu.org/aom.html}.

\section*{Acknowledgements}
The European Pulsar Timing Array (EPTA) is a collaboration of European
institutes to work towards the direct detection of low-frequency GWs
and to implement the Large European Array for Pulsars (LEAP). The
authors acknowledge the support of the colleagues in the EPTA.  We
also like to thank our colleagues A. Deller, M. Kettenis and Z. Paragi
for sharing their knowledge of VLBI and an anonymous referee for
 their comments.  The work reported in this paper has been funded
by the ERC Advanced Grant ``LEAP'', Grant Agreement Number 227947 (PI
M. Kramer). Pulsar astronomy at the University of Manchester and
access to the Lovell telescope are funded through a consolidated grant
from STFC. RS acknowledges support from the European Research Council
under the European Union's Seventh Framework Programme (FP/2007-2013)
/ ERC Grant Agreement n. 617199.

\bibliographystyle{mnras}
\bibliography{pipeline}

\begin{thebibliography}{}
\makeatletter
\relax
\def\mn@urlcharsother{\let\do\@makeother \do\$\do\&\do\#\do\^\do\_\do\%\do\~}
\def\mn@doi{\begingroup\mn@urlcharsother \@ifnextchar [ {\mn@doi@}
  {\mn@doi@[]}}
\def\mn@doi@[#1]#2{\def\@tempa{#1}\ifx\@tempa\@empty \href
  {http://dx.doi.org/#2} {doi:#2}\else \href {http://dx.doi.org/#2} {#1}\fi
  \endgroup}
\def\mn@eprint#1#2{\mn@eprint@#1:#2::\@nil}
\def\mn@eprint@arXiv#1{\href {http://arxiv.org/abs/#1} {{\tt arXiv:#1}}}
\def\mn@eprint@dblp#1{\href {http://dblp.uni-trier.de/rec/bibtex/#1.xml}
  {dblp:#1}}
\def\mn@eprint@#1:#2:#3:#4\@nil{\def\@tempa {#1}\def\@tempb {#2}\def\@tempc
  {#3}\ifx \@tempc \@empty \let \@tempc \@tempb \let \@tempb \@tempa \fi \ifx
  \@tempb \@empty \def\@tempb {arXiv}\fi \@ifundefined
  {mn@eprint@\@tempb}{\@tempb:\@tempc}{\expandafter \expandafter \csname
  mn@eprint@\@tempb\endcsname \expandafter{\@tempc}}}

\bibitem[\protect\citeauthoryear{{Abbott} et~al.,}{{Abbott} et~al.}{2016}]{GW}
{Abbott} B.~P.,  et~al., 2016, Physical Review Letters, 116, 061102

\bibitem[\protect\citeauthoryear{{Bassa} et~al.,}{{Bassa} et~al.}{2016}]{leap}
{Bassa} C.~G.,  et~al., 2016, MNRAS, 456, 2196

\bibitem[\protect\citeauthoryear{{Deller}, {Tingay}, {Bailes}  \&
  {West}}{{Deller} et~al.}{2007}]{difx}
{Deller} A.~T.,  {Tingay} S.~J.,  {Bailes} M.,   {West} C.,  2007, PASP, 119,
  318

\bibitem[\protect\citeauthoryear{{Demorest} et~al.,}{{Demorest}
  et~al.}{2013}]{demorest}
{Demorest} P.~B.,  et~al., 2013, ApJ, 762, 94

\bibitem[\protect\citeauthoryear{{Desvignes} et~al.,}{{Desvignes}
  et~al.}{2016}]{epta}
{Desvignes} G.,  et~al., 2016, MNRAS, 458, 3341

\bibitem[\protect\citeauthoryear{{Detweiler}}{{Detweiler}}{1979}]{Detweiler}
{Detweiler} S.,  1979, ApJ, 234, 1100

\bibitem[\protect\citeauthoryear{{Haehnelt}}{{Haehnelt}}{1994}]{haehnelt}
{Haehnelt} M.~G.,  1994, MNRAS, 269, 199

\bibitem[\protect\citeauthoryear{{Hamaker} \& {Bregman}}{{Hamaker} \&
  {Bregman}}{1996}]{hamaker}
{Hamaker} J.~P.,  {Bregman} J.~D.,  1996, A\&AS, 117, 161

\bibitem[\protect\citeauthoryear{{Hellings} \& {Downs}}{{Hellings} \&
  {Downs}}{1983}]{Hellings}
{Hellings} R.~W.,  {Downs} G.~S.,  1983, ApJ, 265, L39

\bibitem[\protect\citeauthoryear{{Hotan}, {van Straten}  \&
  {Manchester}}{{Hotan} et~al.}{2004}]{psrchive}
{Hotan} A.~W.,  {van Straten} W.,   {Manchester} R.~N.,  2004, Publ. Astron.
  Soc. Australia, 21, 302

\bibitem[\protect\citeauthoryear{{Jaffe} \& {Backer}}{{Jaffe} \&
  {Backer}}{2003}]{jaffe}
{Jaffe} A.~H.,  {Backer} D.~C.,  2003, ApJ, 583, 616

\bibitem[\protect\citeauthoryear{{Karuppusamy}}{{Karuppusamy}}{2011}]{roach}
{Karuppusamy} R.,  2011, in {Burgay} M.,  {D'Amico} N.,  {Esposito} P.,
  {Pellizzoni} A.,   {Possenti} A.,  eds,  American Institute of Physics
  Conference Series Vol. 1357, American Institute of Physics Conference Series.
  pp 89--90

\bibitem[\protect\citeauthoryear{{Karuppusamy}, {Stappers}  \& {van
  Straten}}{{Karuppusamy} et~al.}{2008}]{puma2}
{Karuppusamy} R.,  {Stappers} B.,   {van Straten} W.,  2008, Publ. Astron. Soc.
  Australia, 120, 191

\bibitem[\protect\citeauthoryear{{Keimpema} et~al.,}{{Keimpema}
  et~al.}{2015}]{SFX}
{Keimpema} A.,  et~al., 2015, Experimental Astronomy, 39, 259

\bibitem[\protect\citeauthoryear{{Kramer} \& {Stappers}}{{Kramer} \&
  {Stappers}}{2010}]{leap2010}
{Kramer} M.,  {Stappers} B.,  2010, in ISKAF2010 Science Meeting. p.~34

\bibitem[\protect\citeauthoryear{{Lee}, {Wex}, {Kramer}, {Stappers}, {Bassa},
  {Janssen}, {Karuppusamy}  \& {Smits}}{{Lee} et~al.}{2011}]{single}
{Lee} K.~J.,  {Wex} N.,  {Kramer} M.,  {Stappers} B.~W.,  {Bassa} C.~G.,
  {Janssen} G.~H.,  {Karuppusamy} R.,   {Smits} R.,  2011, MNRAS, 414, 3251

\bibitem[\protect\citeauthoryear{{Lentati} et~al.,}{{Lentati}
  et~al.}{2015}]{lentati}
{Lentati} L.,  et~al., 2015, MNRAS, 453, 2576

\bibitem[\protect\citeauthoryear{{Lommen}}{{Lommen}}{2012}]{lommen}
{Lommen} A.~N.,  2012, Journal of Physics Conference Series, 363, 012029

\bibitem[\protect\citeauthoryear{{Manchester} et~al.,}{{Manchester}
  et~al.}{2013}]{ppta}
{Manchester} R.~N.,  et~al., 2013, Proc. Astr. Soc. Aust., 30, e017

\bibitem[\protect\citeauthoryear{{Manchester} et~al.,}{{Manchester}
  et~al.}{2015}]{tempo}
{Manchester} R.,  et~al., 2015, {Tempo: Pulsar timing data analysis},
  Astrophysics Source Code Library (\mn@eprint {ascl} {1509.002})

\bibitem[\protect\citeauthoryear{{Nita} \& {Gary}}{{Nita} \&
  {Gary}}{2010a}]{nita1}
{Nita} G.~M.,  {Gary} D.~E.,  2010a, PASP, 122, 595

\bibitem[\protect\citeauthoryear{{Nita} \& {Gary}}{{Nita} \&
  {Gary}}{2010b}]{nita2}
{Nita} G.~M.,  {Gary} D.~E.,  2010b, MNRAS, 406, L60

\bibitem[\protect\citeauthoryear{{Ryan} \& {Vandenberg}}{{Ryan} \&
  {Vandenberg}}{1980}]{calc}
{Ryan} J.~W.,  {Vandenberg} N.~R.,  1980, in Bulletin of the American
  Astronomical Society. p.~457

\bibitem[\protect\citeauthoryear{{Sanidas}, {Battye}  \& {Stappers}}{{Sanidas}
  et~al.}{2012}]{cosmic}
{Sanidas} S.~A.,  {Battye} R.~A.,   {Stappers} B.~W.,  2012, Phys. Rev. D, 85,
  122003

\bibitem[\protect\citeauthoryear{{Schwab} \& {Cotton}}{{Schwab} \&
  {Cotton}}{1983}]{schwab}
{Schwab} F.~R.,  {Cotton} W.~D.,  1983, AJ, 88, 688

\bibitem[\protect\citeauthoryear{{Sesana}, {Vecchio}  \& {Colacino}}{{Sesana}
  et~al.}{2008}]{sesana}
{Sesana} A.,  {Vecchio} A.,   {Colacino} C.~N.,  2008, MNRAS, 390, 192

\bibitem[\protect\citeauthoryear{{Shannon} et~al.,}{{Shannon}
  et~al.}{2015}]{shannon}
{Shannon} R.~M.,  et~al., 2015, Science, 349, 1522

\bibitem[\protect\citeauthoryear{{Taylor}}{{Taylor}}{1992}]{template}
{Taylor} J.~H.,  1992, Philosophical Transactions of the Royal Society of
  London Series A, 341, 117

\bibitem[\protect\citeauthoryear{{Taylor} \& {Weisberg}}{{Taylor} \&
  {Weisberg}}{1982}]{Taylor}
{Taylor} J.~H.,  {Weisberg} J.~M.,  1982, ApJ, 253, 908

\bibitem[\protect\citeauthoryear{{The NANOGrav Collaboration} et~al.,}{{The
  NANOGrav Collaboration} et~al.}{2015}]{nanograv}
{The NANOGrav Collaboration} et~al., 2015, ApJ, 813, 65

\bibitem[\protect\citeauthoryear{{van Straten} \& {Bailes}}{{van Straten} \&
  {Bailes}}{2011}]{dspsr}
{van Straten} W.,  {Bailes} M.,  2011, Publ. Astron. Soc. Australia, 28, 1

\bibitem[\protect\citeauthoryear{{van Straten}, {Demorest}  \& {Oslowski}}{{van
  Straten} et~al.}{2012}]{dada}
{van Straten} W.,  {Demorest} P.,   {Oslowski} S.,  2012, Astronomical Research
  and Technology, 9, 237

\makeatother
\end{thebibliography}

\end{document}